\documentclass[a4paper,twocolumn]{article}

\usepackage{amsmath}
\usepackage{graphicx}
\usepackage[noblocks]{authblk}
\usepackage[top=3cm,bottom=2cm]{geometry}
\usepackage{balance}

\begin{document}

\title{\bf SEMI-ANALYTICAL CALCULATIONS FOR PARAMETERS OF BOILING LAYER IN ISENTROPIC EXPANSION OF WARM DENSE MATTER WITH VAN DER WAALS EQUATION OF STATE}
\author[1]{Borovikov\,D.S}
\affil[1]{MIPT, Dolgoprudny, Russia}
\author[2]{ Iosilevskiy\,I.L.}
\affil[2]{JIHT RAS, Moscow, Russia}
\date{}
\maketitle

\textbf{Introduction.}
Properties of isentropic expansion of warm dense matter (WDM), 
which is created by intense energy flux (strong shock compression 
\cite{For-UFN} or intence isochoric heating by laser or/and heavy ion 
beam \cite{[9]}) are often discussed in situation when: 
($i$) -- thermodynamic trajectory of such expansion crosses binodal of liquid-gas phase transition (i.e. boiling curve), and 
($ii$) -- isentropic expansion in two-phase region is going along \emph{equilibrium} branch (not metastable one) of two-phase mixture isentrope. 
Possibility of formation of remarcable "plateau" in density and pressure
profiles 
was studied previously in \cite{Anisimov:1999} (and claimed in 
\cite{Barbel-07}) where the problem of metal surface ablation 
under the action of strong laser radiation was considered.  
Similar plateau was announced also in hydrodynamic symulations 
of equilibrium isentropic expansion with real two-phase EOS of some metals \cite{Barnard-07}. 

Peculiarity of "freezing" of finite portion of expanding matterial in extended and uniform state ("boiling" layer) corresponding to binodal of gas-liquid or/and other phase transitions, -- so called "phase freezeout" -- and prospects of applications of this phenomenon for intended generation of uniform and extended zone of previously unexplored states of matter just at the two-phase bundry were discussed in \cite{Iosilevskiy:2011}.

Features and parameters of such "boiling" liquid layer, which could arise under isentropic expansion of WDM, are studied with the use of simplest
\emph{van der Waals} equation of state (VdW-EOS). 
Possibility of demonstrable and semi-analytical description of thermo- and hydrodynamics of the process is main advantage of this EOS. 
The simplest self-similar 1D hydrodynamics of isentropic expansion of 
semi-infinite VdW-WDM in planar geometry is analyzed. 

\textbf{Caloric EOS in one-phase region.}
It is known that so-called "reduced" caloric EOS in form $\epsilon(P,\rho)$
is the only thermodynamic EOS, which is necessary for description of all 
\emph{adiabatic flows}: i.e. shock compression; isentropic expansion and compression; throttling process 
(Joule--Thompson); adiabatic expansion in vacuum, istant spinodal decomposition etc.

Thermal EOS of van der Waals (VdW) fluid in one-phase region 
is well-known (see (1)). 
In the simplest case of gas of structureless  molecules isochoric heat capacity is constant: $c_V=const$. 
Internal energy and entropy (without insignificant constants) in one-phase region can be calculated as \cite{Landau:2005}:

\begin{equation}
(P+a\rho^2)(1-\rho b)=\rho T,
\label{TEOS}
\end{equation}
\begin{equation}
\epsilon(P,\rho) =c_{V}T-\rho a=c_{V}\frac{1-\rho b}
{\rho}(P+a \rho^2)-\rho a,
\label{1pEnergy}
\end{equation}
\begin{equation}
\label{1pEntropy}
s(P,\rho) =c_{V}\ln (P+a \rho^2)+c_{P}^{id}\ln \frac{1-\rho b}{\rho},
\end{equation}

 where $a$, $b$ -- parameters of VdW EOS and 
$c_{P}^{id}=c_{V}+1=const$.

Isentrope in one-phase region follows from (\eqref{1pEntropy}):

\begin{equation}
P_{s}(\rho)=\text{const}\left(\frac{\rho b}{1-\rho b}\right)^{\gamma_{0}}-a\rho^{2},\label{1pPAd}
\end{equation}
where $\gamma_{0}=c^{id}_P/c_V$.

\textbf{EOS in two-phase region.} For equilibrium states in two-phase region EOS \eqref{TEOS}, \eqref{1pEnergy} and \eqref{1pEntropy} should be corrected with use of Maxwell rule \cite{Landau:2005}. This correction will lead to pressure becoming unambiguous function of temperature $P=P(T)$. From here one can derive form of EOS in two-phase region.
\begin{align}
\left(\frac{\partial\epsilon}{\partial v}\right)_{P}&=\left(\frac{\partial\epsilon}{\partial v}\right)_{T}=T\frac{dP}{dT}-P=f(P),\nonumber\\
\left(\frac{\partial s}{\partial v}\right)_{P}&=\left(\frac{\partial s}{\partial v}\right)_{T}=\frac{dP}{dT}.\label{f appearence}
\end{align}
Using \eqref{f appearence} one obtains:
\begin{align}
\epsilon(P,\rho)&=\frac{f(P)}{\rho}+g(P),\label{2pEnergy}\\
s(P,\rho)&=\frac{1}{\rho}\frac{dP}{dT}+\phi(P),\label{2pEntropy}
\end{align}
where functions $f(P)$, $g(P)$ and $\phi(P)$ can be obtained numerically.
\begin{align}
f(P)&=\left(\frac{\partial\epsilon}{\partial v}\right)_{P}=\frac{\epsilon_{g}-\epsilon_{l}}{v_{g}-v_{l}},\nonumber\\
g(P)&=\left(\frac{\partial\rho\epsilon}{\partial\rho}\right)_{P}=\frac{\rho_{g}\epsilon_{g}-\epsilon_{l}\rho_{l}}{\rho_{g}-\rho_{l}}\label{fg1expr},
\end{align}
where $x_{g}$ and $x_{l}$ - value of physical quantity $x$ on gaseous and liquid border of phase transition correspondingly.

It should be stressed that equations \eqref{2pEnergy} and \eqref{2pEntropy} are universal forms of EOS in two-phase region for all phase transitions of $1^{st}$ order: they are derived without referrence to particular EOS. 

Taking into account that $\left(\frac{\partial\epsilon}{\partial P}\right)_{v}=T\left(\frac{\partial s}{\partial P}\right)_{v}$ one can derive universal dependence between functions $g(P)$ and $\phi(P)$:
\begin{equation}
\phi'(P)=\frac{g'(P)}{T}.\label{PhiGBond}
\end{equation}
For values of all quantities should coincide on border between one- and two-phase regions -- binodal, -- another expressions and mutual dependence for $f(P)$ and $g(P)$ can be derived with using \eqref{fg1expr} and EOS \eqref{1pEnergy} :
\begin{align}
f(P)&=a\rho_{l}\rho_{g},\nonumber\\
g(P)&=c_{V}T-a(\rho_{l}+\rho_{g}),\nonumber\\
g(P)&=c_{V}T-\frac{a}{b}\left(1-\frac{P}{f(P)}\right)\label{fg2expr}
\end{align}

In two-phase region Poisson adiabat can be obtained numerically:
\begin{equation}
\left(\frac{\partial P}{\partial v}\right)_{s}=-\frac{\left(\frac{\partial\epsilon}{\partial v}\right)_P+P}{\left(\frac{\partial\epsilon}{\partial P}\right)_v}=-\frac{f(P)+P}{f'(P)v+g'(P)}.\label{2pPAd}
\end{equation}

\textbf{Advantages of van der Waals EOS.} The main advantage of VdW EOS is that simple expression for $f(P)$ and $g(P)$ and analytical dependence \eqref{fg2expr} between them can be obtained. It also means, that knowledge of dependence $P(T)$ in two-phase region is sufficient to describe state of fluid in it.

Advantage of all EOS's with two parameters is that all quantities can be transformed into ones without dimension: reduced form of EOS can be used. It also can be applied to functions $f(P)$ and $g(P)$. Since there are characteristic values for pressure and density -- $P_{crit}$ and $\rho_{crit}$ -- one can derive characteristic values for length, time and, what is especially important for this investigation, for velocity:
\begin{equation}
u_{ch}=\sqrt{\frac{P_{crit}}{\rho_{crit}}}.\nonumber
\end{equation}
Therefore hydrodynamics of van der Waals fluid can be described in nondimensional form.

\textbf{Hydrodynamics in planar geometry.} Process of isentropic expansion of semi-infinite layer in vacuum is considered. It is described by equations of hydrodynamics and EOS\cite{Landau:2006, Zeldovich:2005}:
\begin{align}
&\frac{\partial\rho}{\partial t}+\frac{\partial\rho u}{\partial x}=0;\nonumber\\
&\frac{\partial u}{\partial t}+u\frac{\partial u}{\partial  x}=-\frac{1}{\rho}\frac{\partial P}{\partial x};\nonumber\\
&\frac{\partial\epsilon}{\partial t}+u\frac{\partial\epsilon}{\partial x}=-\frac{P}{\rho}\frac{\partial u}{\partial x};\nonumber\\
&\epsilon=\epsilon(P,\rho).\label{HydEq}
\end{align}
In semi-infinite planar geometry the solution is self-similiar: all quantities depend on variable $\xi=x/t$ only  \cite{Landau:2006, Zeldovich:2005}. Taking that into account one can transform system \eqref{HydEq}:
\begin{align}
\xi&=u\pm c;\nonumber\\
u&=\pm\int c\;\frac{\mathrm{d}\rho}{\rho};\nonumber\\
P&=P_{s}(\rho),\label{HydEqTr}
\end{align}
where $c=\sqrt{\left(\frac{\partial P}{\partial\rho}\right)_{s}}$ -- speed of sound, sign (plus or minus) is determined by direction of material flow of material. In \eqref{HydEqTr} we also took into account the fact, that in our case flow is isentropic. So we substituted EOS for Poisson adiabat $P=P_{s}(\rho)$.

\begin{figure}[!ht]
  \includegraphics[width=0.95\columnwidth]{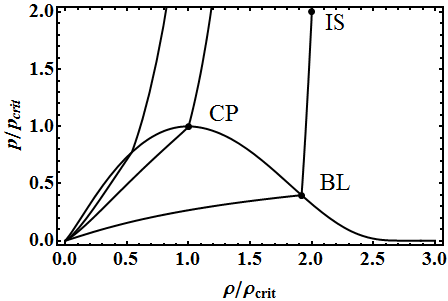}
  \caption{Interception of Poisson adiabats with binodal. For VdW fluid $\rho\leq 3\rho_{crit}$}
  \label{fig1}
\end{figure}

\textbf{"Phase freezeout".} It is clearly seen on \mbox{Figure \ref{fig1}}, that adiabat has a breakpoint on interception with binodal. It means, that sound of speed has a jump at this point. For mass velocity is continuous function (in other case the law of flow conservation would be violated), we also have jump of variable $\xi$ from value $\xi^+_b$ -- right before entering two-phase region -- to value $\xi^-_b$ -- right after it.

For interception point is unambiguously determined by choice of adiabat, value of sound of speed jump is determined by entropy of process.

From here we can conclude, that for all values  \mbox{$\xi^-_b<\xi<\xi^+_b$} all hydro- and thermodynamical parameters stay constant -- extensive uniform "boiling" layer emerges.

\begin{figure}[h]
\centering
    \includegraphics[width=0.95\columnwidth]{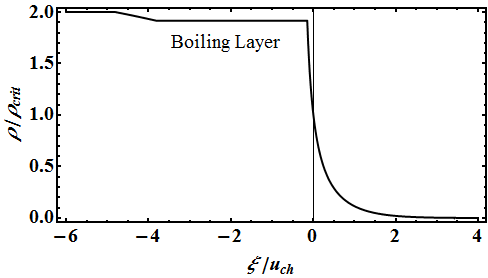}
    \includegraphics[width=0.95\columnwidth]{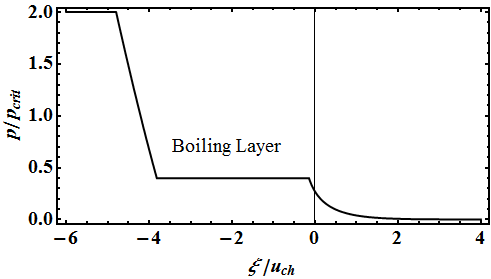}
    \caption{Example of flow with "boiling" layer.}    
    \label{fig2}
\end{figure}

Example of the problem`s solution with clearly seen "boiling" layer is depicted on Figure \ref{fig2}. Initial state and "boiling" layer for the example are depicted as points IS and BL correspondingly on Figure \ref{fig1}.

One of the parameters of phenomenon is part of disturbed matter, confined in "boiling" layer $\alpha$:
\begin{equation}
\alpha=\frac{\rho_b}{\rho_0}\frac{c^+_b-c^-_b}{c_0},\label{alpha}
\end{equation}
where $\rho_0$ and $c_0$ -- density and speed of sound in initial state, $\rho_b$ -- density on "boiling" layer, $c^+_b$ and $c^-_b$ -- sound of speed before and after entering two-phase region.

For example, for the case, depicted on Figure \ref{fig2}, $\alpha=73.3\%$.

Parameters $\alpha$ and $u_b$ -- mass velocity on "boiling" layer -- are determined by choice of initial state. If entropy is already chosen, we can say, that they determined by value of initial pressure.

So we identified, that the phenomenon has two degrees of freedom: entropy and initial pressure.

\begin{figure}[!t]
  \includegraphics[width=0.95\columnwidth]{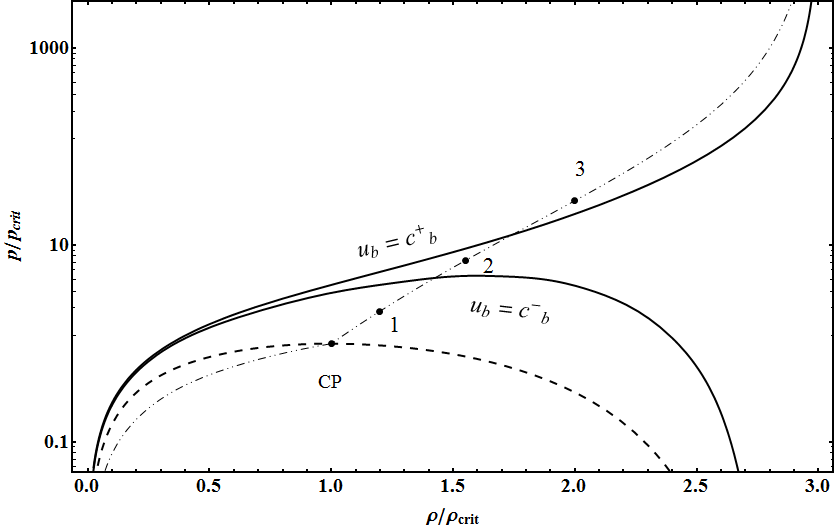}
  \caption{Regimes of "phase freezeout".}
  \label{fig3}
\end{figure}

\begin{figure}[t]
    \includegraphics[width=0.95\columnwidth]{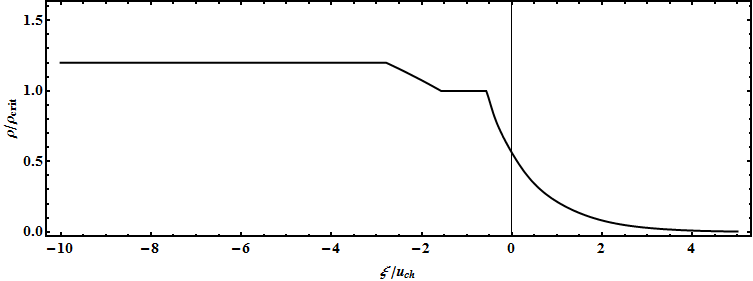}
    \includegraphics[width=0.95\columnwidth]{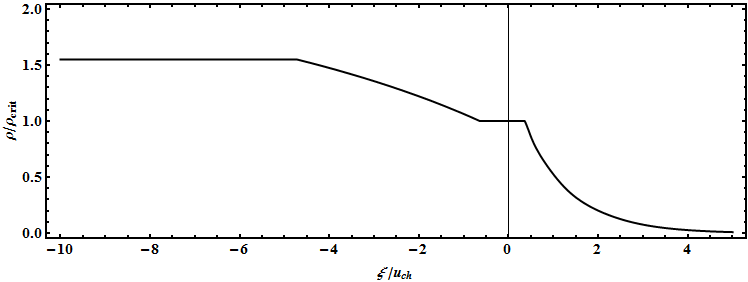}
    \includegraphics[width=0.95\columnwidth]{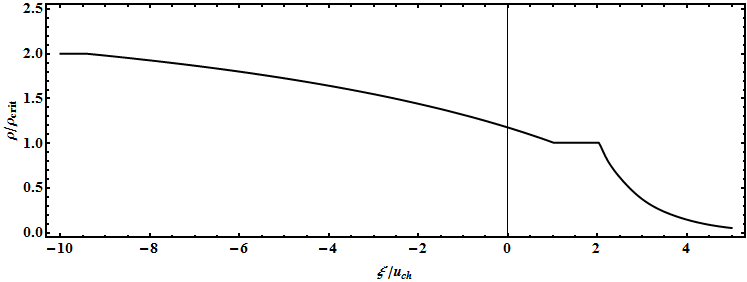}
    \caption{Density profiles in different regimes of "phase freezeout".}    
    \label{fig4}
\end{figure}

\textbf{Regimes of "phase freezeout".} It was obtained, that "boiling" layer displays different behaviors: direction of expansion of the layer changes with initial conditions. There are three regimes of phenomenon:
\begin{enumerate}
\item $c_b^+>c_b^->u_b$ -- the whole layer moves in direction, opposite to vacuum,
\item $c_b^+>u_b>c_b^-$ -- the layer expands in both directions,
\item $u_b>c_b^+>c_b^-$ -- the whole layer moves toward vacuum.
\end{enumerate}

On Figure \ref{fig3} initial states of boundary regimes $u_b=c_b^-$ and $u_b=c_b^+$ are depicted as thick lines. Dashed line is binodal, doted line is critical adiabat.

On Figure \ref{fig4} different regimes are illustrated. Initial states for all three cases are depicted as points 1, 2 and 3 on Figure \ref{fig3}. For  the \mbox{case 1} $\alpha=29.9\%$, for the \mbox{case 2} $\alpha=13.7\%$, for the \mbox{case 3} $\alpha=5.3\%$.

\textbf{Conclusion.} Features of the discussed regimes of isentropic expansion of WDM-material makes it possible to discuss, at least formally, possibility of generation of uniform and extensive sample of investigated substance, which we are interested in, just on the boiling boundary of liquid-gas phase transition (including the region of its critical point). Size of this zone is determined by interception point of binodal and Poisson adiabat and initial pressure. It grows with growth of sound of speed jump at binodal, which is higher at higher densities, and decreases with growth of initial pressure.

\textbf{Acknowledgements.} The work was supported by MIPT Education Center "Physics of high energy density matter" and by RAS Scientific Program "Physics of extreme states of matter" and partially (I.L.I.) by Extreme Matter Institute-EMMI (Germany).\balance

\end{document}